\documentclass[12pt,preprint]{aastex}
\def\bmath#1{\mbox{\boldmath$#1$}}

\slugcomment{Submitted to the Astrophysical Journal}

\shorttitle{Direction of Outflow} \shortauthors{Matsumoto and
Tomisaka}

\begin{document}

\title{Directions of Outflows, Disks, Magnetic Fields, and Rotation of YSOs in Collapsing Molecular Cloud Cores}

\author{Tomoaki Matsumoto\altaffilmark{1}} \email{matsu@i.hosei.ac.jp}

\and

\author{Kohji Tomisaka\altaffilmark{2}} \email{tomisaka@th.nao.ac.jp}

\altaffiltext{1}{Department of Humanity and Environment, Hosei
 University, Fujimi, Chiyoda-ku, Tokyo 102-8160, Japan, also Visiting
 Professor of National Astronomical Observatory}
 \altaffiltext{2}{Division of Theoretical Astrophysics, National
 Astronomical Observatory, Mitaka, Tokyo 181-8588, Japan}

\begin{abstract}

The collapse of slowly rotating molecular cloud cores threaded by
magnetic fields is investigated by high-resolution numerical
simulation. Outflow formation in the collapsing cloud cores is also
followed. In the models examined, the cloud core and parent cloud
rotate rigidly and are initially threaded by a uniform magnetic
field. The simulations show that the cloud core collapses along the
magnetic field lines. The magnetic field in the dense region of the
cloud core rotates faster than that of the parent cloud as a
consequence of spin-up of the central region during the collapse. The
cloud core exhibits significant precession of the rotation axis,
magnetic field, and disk orientation, with precession highest in the
models with low initial field strength ($\lesssim 20 \, \mu {\rm
G}$). Precession in models with initial fields of $\sim 40\,\mu {\rm
G}$ is suppressed by strong magnetic braking. Magnetic braking
transfers angular momentum form the central region and acts more
strongly on the component of angular momentum oriented perpendicular
to the magnetic field. After the formation of an adiabatic core,
outflow is ejected along the local magnetic field lines. Strong
magnetic braking associated with the outflow causes the direction of
angular momentum to converge with that of the local magnetic field,
resulting in the convergence of the local magnetic field, angular
momentum, outflow, and disk orientation by the outflow formation
phase. The magnetic field of a young star is inclined at an angle of
no more than $30^\circ$ from that of the parent cloud at initial field
strengths of $\sim 20 \, \mu {\rm G}$, while at an initial field
strength of $\sim 40\,\mu {\rm G}$, the magnetic field of the young
star is well aligned with that of the parent cloud.

\end{abstract}


\section{Introduction}

Magnetic fields are believed to play an important role in the
gravitational collapse of molecular cloud cores. The characteristics
of the outflow and jet associated with young stars are related to the
magnetic field and the rotation of the (proto)star and circumstellar
disk. Observations of polarization have suggested that the outflows
and jets often tend to be aligned with each other and with the
cloud-scale magnetic field
\citep*[e.g.,][]{Cohen84,Strom86,Vrba86,Vrba88,Tamura89,Jones03}. The
polarization of young stellar objects (YSOs) also suggest that
circumstellar dust disks around young stars are aligned perpendicular
to the magnetic field \citep[e.g.,][]{Moneti84,Tamura89}. Recent
high-resolution observations of submillimeter polarization have
resolved the magnetic fields around young stars on the $\sim 1000$~AU
scale, which is comparable to the outflow scale, and interactions
between the magnetic fields, the outflows, and the envelopes have been
inferred \citep*{Momose01,Henning01,Wolf03,Vallee03}. \citet{Wolf03}
suggested that two Bok globules are associated with outflows parallel
to the magnetic field, while the two other globules are associated
with outflows perpendicular to the magnetic field. Thus, the findings
for the correlation between outflow, envelope, and magnetic field have
yet to converge.

Although the alignment of outflows and jets with the magnetic field
has been discussed intensively in observational studies, very little
theoretical consideration has been given to this problem. Theoretical
studies of the collapse of a magnetized cloud have been restricted to
models of aligned rotators, in which the rotation axis is parallel to
the magnetic field
\citep*[e.g.,][]{Nakamura95,Tomisaka96,Nakamura99}. 
Several other
studies \citep{Mouschovias79,Dorfi82,Dorfi89} have investigated
inclined rotators. However, a collapsing inclined rotator has not been
investigated for at least 15 years. 

It is believed that ambipolar diffusion may play an important
role in the gravitational collapse of cloud core.  Theoretical studies
including the ambipolar diffusion have done by Mouschovias and the
co-authors \citep[e.g.,][]{Fiedler93,Ciolek94}, and recently by
\citet{Nakamura03}.  However, they simulated collapse of
aligned rotators by two-dimensional simulations. 
The three-dimensional simulations are necessary to investigate 
alignment of an outflow with magnetic field in a collapsing inclined
rotator. 
At this moment, three-dimensional simulations including ambipolar
diffusion just begin by \citet{Hosking04} using a SPH scheme.
In order to investigate collapse of inclined rotators, we
perform three-dimensional simulations assuming the ideal MHD.
\citet{Nakano02} shows that timescale of magnetic flux loss
is considerably longer than the free-fall time in $n \lesssim
10^{11}\,{\rm cm}^{-3}$,  indicating that the 
ideal MHD is valid during the early phase of dynamical collapse.

Observations suggest that the
rotation axis of the molecular cloud is not generally aligned with the
magnetic field, as exemplified by the filamentary molecular cloud of
Taurus molecular cloud 1 (TMC1), where the magnetic field is
perpendicular to the filamentary axis \citep{Tamura89} and a velocity
gradient perpendicular to the filamentary axis has also been observed
for TMC1 \citep{Olano88}. Accordingly, molecular cloud cores, as
substructures of molecular clouds, probably have inclined rotational
axes with respect to the cloud-scale magnetic field. 
\citet{Goodman93} indicates that the orientation of rotation axes of
cloud cores distribute randomly in the Taurus region, and suggests that
the rotation axes are uncorrelated with the cloud-scale magnetic field.


Simulations of inclined rotators have also been performed, but with
the objective of studying magnetic braking of molecular clouds rather
than molecular cloud cores \citep{Dorfi82,Dorfi89}. In these
simulations, the magnetic field is artificially anchored to the
boundary of the computation box, and only the cloud of interest
rotates, while the surrounding gas rests at the initial state. The
latter assumption induces strong shear flow at the cloud boundary,
which promotes magnetic braking. Furthermore, the initial state also
requires some explanation. If a cloud core forms from a turbulent
cloud, both the cloud core and the envelope must rotate. In the
present study, realistic models in which both the cloud core and
parent cloud rotate and which are initially threaded by a uniform
magnetic field are considered.  The collapse of the cloud core in
these models produces differential rotation even from a uniformly
rotating core-envelope system, which promotes magnetic braking.

Theoretical studies of outflow formation in the collapsing core have
also been restricted to axisymmetric models
\citep*{Tomisaka98,Tomisaka02,Allen03}. Axisymmetric models are
applicable only if the rotation axis and magnetic field converge to be
parallel, related to magnetic braking before outflow formation. As is
shown by the simulations in this study, convergence cannot be
completed before outflow formation. The effects of outflow on
convergence are considered here in some detail.


In this paper, the gravitational collapse of slowly rotating
magnetized cloud cores threaded by inclined magnetic fields is
investigated. In \S~\ref{sec:basiceq} and \S~\ref{sec:model}, the
basic equations governing the simulations and the model of magnetized
molecular cloud cores are introduced. In \S~\ref{sec:method}, the
methods of numerical simulations are presented, and in
\S~\ref{sec:results}, the results of the simulations are shown and the
inclination of the magnetic field, angular momentum, disk, and
outflows are estimated. In \S~\ref{sec:discussion}, precession and
alignment of several axes of a YSO are discussed with respect to
magnetic braking, and outflow velocities and momenta are
considered. Finally, the paper is concluded in \S~\ref{sec:summary}.

\section{Basic Equations}
\label{sec:basiceq}

Self-gravitational magnetohydrodynamic (MHD) equations are assumed, as
described by
\begin{equation}
\frac{\partial \rho}{\partial t} + \nabla\cdot\left( \rho
\bmath{v}\right) = 0 \;,
\label{eq:mass}
\end{equation}
\begin{equation}
\frac{\partial}{\partial t}(\rho \bmath{v}) +\nabla\cdot(\rho
\bmath{v} \bmath{v}^{T}) +\nabla \left(P + \frac{|\bmath{B}|^2}{8\pi}
\right) - \nabla \cdot \left( \frac{\bmath{B}\bmath{B}^{T}}{4\pi}
\right) + \rho \nabla \Psi = 0 \;,
\label{eq:momentum}
\end{equation}
\begin{equation}
\frac{\partial}{\partial t}\bmath{B} + \nabla \cdot (\bmath{v}
\bmath{B}^{T} - \bmath{B} \bmath{v}^{T} )=0\;,
\label{eq:induction}
\end{equation}
and
\begin{equation}
\nabla^2 \Psi = 4 \pi G \rho \;,
\label{eq:poisson}
\end{equation}
where $\rho$, $\bmath{v}$, $P$, $\bmath{B}$, $\Psi$, and $G$ denote
density, velocity, pressure, magnetic field, gravitational potential,
and the gravitational constant. 
The transpose of $\bmath{B}$ is denoted by $\bmath{B}^{T}$.
The equation of state for a molecular
cloud is well described by an isothermal gas for low-density regions
and a polytropic gas for high-density regions \citep{Tohline82}. Two
types of equation of state are therefore employed, i.e.,
\begin{equation}
P = \left\{
\begin{array}{ll}
c_s^2 \rho & {\rm for}\; \rho < \rho_{\rm cr}\\ c_s^2 \rho_{\rm cr}
(\rho/\rho_{\rm cr})^{5/3} & {\rm for}\; \rho \geq \rho_{\rm cr}
\end{array}
\right. \;,
\label{eq:pressure}
\end{equation}
where $c_s$ and $\rho_{\rm cr}$ denote the isothermal sound speed and
the critical density. The gas temperature is assumed to be 10~K ($ c_s
= 0.19 \,{\rm km}\,{s}^{-1}$) below the critical density $\rho_{\rm
cr} = 2 \times 10^{-13}\,{\rm g}\,{\rm cm}^{-3}$ ($ n _{\rm cr} =
5.24\times10^{10}\,{\rm cm}^{-3} $), and to increase adiabatically
above it.

\section{Models of Cloud Cores}
\label{sec:model}

As an initial model of a molecular cloud core, we consider a slowly
rotating, spherical, isothermal cloud threaded by a uniform magnetic
field. The cloud is confined by a uniform ambient gas that mimics a
parent molecular cloud. Both the cloud and the ambient gas have a
uniform magnetic field and rotate with the same angular rotation
speed.

As a template for a molecular cloud core, we consider the density
profile of the critical Bonnor-Ebert sphere
\citep{Ebert1955,Bonnor1956}. If $\varrho_{\rm BE}(\xi)$ denotes the
non-dimensional density profile of the critical Bonnor-Ebert sphere
\citep[see,][]{Chandrasekhar39}, the initial density distribution is
given by
\begin{equation}
\rho(r) = \left\{
\begin{array}{ll}
\rho_0 \varrho_{\rm BE}(r/a) & {\rm for}\; r < R_c\\ \rho_0
\varrho_{\rm BE}(R_c/a) & {\rm for}\; r \geq R_c
\end{array}
\right. \;,
\end{equation}
and
\begin{equation}
a = c_s \left( \frac{f}{4 \pi G \rho_0} \right)^{1/2} \;,
\end{equation}
where $r$, $f$, and $\rho_0$ denote the radius, density enhancement
factor, the initial central density, respectively. The critical
Bonnor-Ebert sphere is obtained when $f=1$. The radius of the cloud is
defined by $R_c = 6.45 a$, where the numerical factor comes from the
non-dimensional radius of the critical Bonnor-Ebert sphere. The
density contrast of the initial cloud is $\rho(0)/\rho(R_c) = 14$.

In this paper, $\rho_0$ is set equal to $1\times10^{-19}\,{\rm
g}\,{\rm cm}^{-3}$, which corresponds to a number density of $n_0 =
2.61 \times 10^4\,{\rm cm}^{-3}$ for the assumed mean molecular weight
of 2.3.  The initial freefall timescale at the center of the cloud is
thus $t_{\rm ff} \equiv (3 \pi / 32 G \rho_0)^{1/2} =
2.10\times10^5\,{\rm yr} $. The density enhancement factor is set at
$f = 1.68$ to allow the rotating magnetized cloud to collapse. The
radius of the cloud is thus $R_c = 0.178$~pc, and the mass of the
cloud is $6.130\,M_\odot$. The parameter $\alpha_0 (\equiv E_{\rm
th}/|E_{\rm grav}|)$ is equal to 0.5, where $E_{\rm th}$ is the
thermal energy of the cloud and $E_{\rm grav}$ is the gravitational
energy.

The initial cloud and ambient gas rotate rigidly with angular velocity
of $\Omega_0$ and rotation axis parallel to the $z$ axis. We assume $
\Omega_0 =7.11\times10^{-7}\,{\rm yr}^{-1} (= 1.50 t_{\rm ff}^{-1})$,
which corresponds to $\beta_0 = 0.02$, where $\beta_0 \equiv E_{\rm
rot}/ |E_{\rm grav}|$ and $E_{\rm rot}$ is the rotation energy. This
value of $\beta_0$ is typical for rotating molecular cloud cores
\citep{Goodman93}.

The initial magnetic field is given by
\begin{equation}
\left(
\begin{array}{c}
B_x \\ B_y \\ B_z
\end{array}
\right) = B_0\left(
\begin{array}{c}
\sin \theta \\ 0 \\ \cos \theta
\end{array}
\right)
\end{equation}
in Cartesian coordinates $(x,y,z)$, where $B_0$ denotes the initial
magnetic field strength and $\theta$ denotes the angle between the
initial magnetic field and the initial angular momentum. When $\theta
= 0$, the magnetic field is aligned with the angular momentum. Various
models can be constructed by changing the model parameters $B_0$ and
$\theta$, allowing the effect of strength and configuration of the
magnetic field on the collapsing cloud cores to be investigated.

Models with four different magnetic field strengths were constructed,
as shown in Table~\ref{table:mag}. The model names are derived from
the initial strength of the magnetic field, strong (SF), moderate (MF)
or weak (WF), and the initial angle $\theta$ (00-90), with the
no-field model denoted NM. The initial strength of a magnetic field is
characterized by three non-dimensional quantities, 
$B_0/B_{\rm cr}$, $E_{\rm mag}/|E_{\rm grav}|$, and $\beta_p$.  
The first quantity $B_0/B_{\rm cr}$ is the ratio of field strength to critical
field strength, where the critical field strength is defined by
$B_{\rm cr} = G^{1/2} \Sigma / 2 \pi$ and $\Sigma$ is the central
column density of the cloud given by $\Sigma = 2 \int_0^{R_c} \rho(r)
dr$. When the initial magnetic field is weaker than the critical
magnetic field, the cloud is magnetically super-critical, leading to
collapse as the gravity exceeds the magnetic pressure and tension
without sufficient thermal pressure support. This paper considers the
case of super-critical clouds. The second quantity $E_{\rm
mag}/|E_{\rm grav}|$ is the ratio of magnetic energy to gravitational
energy, where $E_{\rm mag} = \int B_0^2 / 8 \pi
d\bmath{r}$. Equipartition of magnetic energy and gravitational energy
is achieved ($E_{\rm mag}/|E_{\rm grav}| = 1$) when $ B_0 = 21.9 \,
\mu {\rm G}$ in the present model. All models except for the weak
field models ($B_0 = 7.42 \,\mu {\rm G}$) and model NM have magnetic
energy comparable to the gravitational energy. The third quantity
$\beta_p$ is the plasma beta at the center of the cloud, as defined by
$\beta_p = 8 \pi c_s^2 \rho_0 / B_0^2$, which takes values of the
order of unity except in the weak field models and model NM.

\section{Numerical Methods}
\label{sec:method}

A self-gravitating MHD simulation code was developed by extending the
simulation code of \citet{Matsumoto03b}. The MHD equations and Poisson
equation are solved by a finite difference method with second-order
accuracy in space and time. A nested grid is employed to solve the
central region with higher spatial resolution. The nested grid
consists of a concentric hierarchical Cartesian grid
\citep{Matsumoto03b,Machida04}, and the cell width of each grid
decreases successively by a factor of 2. The coarsest grid is labeled
level $l = 1$, and the $l$th level grid has $2^{l-1}$ times higher
spatial resolution than the coarsest grid. The sub-grids are
automatically formed when $\lambda_J/16$ becomes smaller than the cell
width of the finest grid, where $\lambda_J$ denotes the Jeans length
\citep[c.f., Jeans condition of][]{Truelove97}. The maximum level of
the grid is $l_{\rm max} = 14$ in typical models.  The detail
implementation is shown in \citet{Matsumoto04}.

The numerical fluxes of the MHD equations (\ref{eq:mass}) --
(\ref{eq:induction}) are obtained using the linearized Riemann solver
reported in \citet{Fukuda00}. This solver is based on the method in
\citet{Roe81} but has been extended for solving ideal, barotropic MHD
equations. A MUSCL approach and predictor-corrector method are adopted
here for integration over time in order to achieve second-order
accuracy in space and time \citep[e.g.,][]{Hirsch1990}. Similar to
standard adaptive mesh refinement (AMR), a multi-timestep scheme is
adopted here \citep*[e.g.,][]{chiang92}.
When the calculation of a fine grid is synchronized with that for
coarser grids, the numerical fluxes of the coarse grid at the grid
interfaces (between fine and coarse grids) are modified by the fluxes
calculated by the fine grid in order to conserve the numerical fluxes
at the grid interfaces. 
In order to keep $\nabla \cdot \bmath{B} = 0$ within a round-off error,
the divergence-free staggered mesh of
\citet{Balsara01} and \citet{Balsara99} is employed. 
This method is
similar to the constrained transport approach of \citet{Evans02}, but
has been optimized for the Godunov-type Riemann solver and
hierarchical grids. This method makes it possible to stably solve
even extremely low-$\beta$ regions of $\beta_p = 10^{-5} - 10^{-3}$.

Self-gravity is updated in every step across all grid levels by
solving the Poisson equation (\ref{eq:poisson}) using the multi-grid
iteration for a nested grid developed by \citet{Matsumoto03a}. In this
code, a full multi-grid scheme is employed \citep[see the algorithm
shown in][]{Press91} to accelerate the convergence of the red-black
Gauss-Seidel iteration.

To test the MHD code, we computed MHD shock tube problems introduced by
\citet{Brio88}, confirming that every wave propergates 
across  the grid interfaces without reflection.
Several tests for self-gravity are described in \citet{Matsumoto03a}. 


A boundary condition is imposed at $r=8R_c$, where the magnetic field
and ambient gas rotate at an angular velocity of $\Omega_0$. The
velocity and magnetic field at the boundary are therefore given by
\begin{equation}
\bmath{v}_{\rm b}(t) = \Omega_0 \hat{\bmath{z}} \times \bmath{R} \;,
\end{equation}
and
\begin{equation}
\bmath{B}_{\rm b}(t) = \bmath{B}_0 \cos(\Omega_0 t) + \hat{\bmath{z}}
(\hat{\bmath{z}}\cdot\bmath{B}_0) \left[ 1-\cos(\Omega_0 t) \right] +
\bmath{B}_0 \times \hat{\bmath{z}}\sin(\Omega_0 t) \;,
\end{equation}
where $\hat{\bmath{z}}$ denotes the unit vector in the $z$ direction
and $\bmath{R}$ denotes a position vector of the cylindrical
radius. The density is fixed whenever equal to the initial density of
$\rho(R_c)$ at this radius. The large radius of this boundary
condition eliminates the effect of artificial magnetic braking due to
the boundary condition. Even for a strong field model of $B_0=37.1
~\mu {\rm G}$, the crossing time for an Alfv\'en wave from the center
to the boundary is $1.1 \times 10^6$~yr, which is 5.3 times longer
than the freefall time. Gas is considered to be the only source of
self-gravity in $r \leq R_c$ when solving the Poisson
equation~(\ref{eq:poisson}). Point-symmetry with respect to the origin
is imposed in order to reduce computational cost.

Each grid consists of $ 128 \times 64 \times 128 $ cubic cells in an
$(x,y,z)$ coordinate frame. At the initial stage, five sub-grids
($l_{\rm max} = 5$) are set, where two fine sub-grids contain the
cloud and the remaining three (sub-)grids contain the ambient gas.
The cloud with radius $R_c$ is therefore solved with a grid of $4 \le
l \le l_{\rm max} $, and the cloud is resolved by $ 128 \times 64
\times 128 \times (l_{\rm max}-3)$ cells.

\section{Results}
\label{sec:results}

\subsection{Clouds Threaded by Moderately Strong Magnetic Fields}
\label{sec:moderate}

\subsubsection{Model MF45}
\label{sec:MF45}

Figure~\ref{model4of45slice.eps} shows the evolution of model MF45
($\theta = 45^\circ$) as a series of cross sections. As shown in
Figure~\ref{model4of45slice.eps}{\it f}, the plane of each cross
section is defined so as to include the $z$ and $x$ axes for planes A
and B, oriented parallel to the mean magnetic field in the central
dense region of $\rho \ge 0.1 \rho_{\rm max}$, where $\rho_{\rm max}$
denotes the maximum density. The mean magnetic field in the central
dense region (the central magnetic field) is defined by
\begin{equation}
\bmath{B}_c = \frac{1}{V}\int_{\rho \ge 0.1 \rho_{\rm max}}
\bmath{B}(\bmath{r}) d\bmath{r} \;,
\label{eq:Bc}
\end{equation}
where $V = \int_{\rho \ge 0.1 \rho_{\rm max}}d\bmath{r} $ denotes the
volume of $\rho \ge 0.1 \rho_{\rm max}$. Plane A forms an angle
$\theta_z$ with the $x-z$ plane, and plane B forms an angle $\theta_x$
with the $x-y$ plane, where $\theta_z \equiv \arctan(B_{c,y}/B_{c,x})$
and $\theta_x \equiv \arctan(B_{c,z}/B_{c,y})$.

Figure~\ref{model4of45slice.eps}{\it a} shows the structure of the
entire cloud core in terms of the cross section of plane A at
$\rho_{\max} = 5.58 \times 10^{-18}~{\rm g}~{\rm cm}^{-3}$. The cloud
undergoes runaway collapse, and the central density increases. The
cloud has an oblate density structure with major axis extending almost
perpendicular to the local magnetic field, as the gas falls along the
field lines. The cloud becomes thinner in the denser region, and
collapses with supersonic flow in the direction perpendicular to the
major axis. The central magnetic field of the dense region rotates
around the $z$ axis by up to $\theta_z=32.7^\circ$, while the initial
magnetic field is parallel to the $x-z$ plane ($\theta_z=0$).

Figure~\ref{model4of45slice.eps}{\it b} shows the central cloud at
$\rho_{\rm max}=1.70\times10^{-14}~{\rm g}~{\rm cm}^{-3}$. The cloud
forms a collapsing isothermal disk at the center, oriented
perpendicular to the local magnetic field. The central magnetic field
rotates by up to $\theta_z = 37.7^\circ$ by this stage. This
collapsing magnetized disk resembles that in axisymmetric simulations
by \citet{Nakamura95} and \citet{Tomisaka96}.

When the central density exceeds the critical density $\rho_{\rm cr}$,
an adiabatic core forms at the center of the disk. The adiabatic core
terminates the runaway collapse, and the accretion phase begins, in
which the adiabatic core accretes gas from the disk-like infalling
envelope, as shown in Figure~\ref{model4of45slice.eps}{\it c}. The
adiabatic core has a radius of 7~AU and corresponds to the first core
of \citet{Larson69}.

The core spins faster than the infalling envelope, causing the
magnetic field to become twisted. A bipolar outflow then forms
approximately 400~yr after formation of the adiabatic core. The
evolution of this outflow was followed for $\sim$ 600~yr.
By the final stage, the Alfv\'en velocity increases up to
189~km~s$^{-1}$ ($= 945 c_s$),
which restricted the timestep of the simulation due the Courant
condition and rendered it necessary to
halt the simulations at this stage.
Figures~\ref{model4of45slice.eps}{\it d} and
\ref{model4of45slice.eps}{\it e} show different cross sections at the
final stage of the simulation.  The outflow can be seen to have
extended out to $\simeq~200$~AU in the final stage, remaining anchored
to a small disk of radius $r\simeq 20$~AU around the adiabatic
core. The outflow is perpendicular to the infalling disk and has an
hourglass structure. Gas is ejected from the small disk, but gas also
continues to fall into the adiabatic core through narrow channels near
the polar axes. The field lines are considerably twisted in the
outflow region, as seen in Figures~\ref{3dmodel4.eps}{\it c} and
\ref{3dmodel4.eps}{\it d}. The outflow has a slightly asymmetric
structure with respect to the disk axis. The high-velocity gas has a
spiral structure that traces the magnetic field lines, resembling that
of \citet{Tomisaka98,Tomisaka02} except for the non-axisymmetric
structure. The outflow in this case is categorized as a U-shaped
outflow according to \citet{Tomisaka02}.

Figure~\ref{3dmodel4.eps} shows three-dimensional views of the cloud
core at the final stage at three different scales. In
Figure~\ref{3dmodel4.eps}{\it a}, which shows the entire region of the
cloud core at the same scale as Figure~\ref{model4of45slice.eps}{\it
a}, the disk of the cloud core is clearly perpendicular to the
magnetic field, which is anchored to the outer boundary and is rotated
around the $z$ axis by $19^\circ$. The central disk axis, however, is
rotated by $\simeq 60^\circ$ due to spin-up of the central region, as
seen in Figure~\ref{3dmodel4.eps}{\it c}, but the disk remains
perpendicular to the magnetic field.

The spin-up of the central cloud can be investigated by evaluating the
angular velocity of the cloud center as follows.
\begin{equation}
\Omega_c = \frac{ | \bmath{J}_c | }{M (a_1^2+a_2^2)} \;,
\end{equation}
where the central angular momentum is defined as
\begin{equation}
\bmath{J}_c = \int_{\rho \ge 0.1 \rho_{\rm max}} \bmath{r}\times
\bmath{v} \rho d\bmath{r} \;,
\label{eq:Jc}
\end{equation}
and the mass is defined as
\begin{equation}
M = \int_{\rho \ge 0.1 \rho_{\rm max}} \rho d\bmath{r} \;.
\label{eq:M}
\end{equation}
Here, $a_n$ ($n=1,2,3$) denote the axis lengths of the dense region,
which can be evaluated by measuring the specific moment of inertia,
i.e.,
\begin{equation}
I_{ij} = \frac{1}{M} \int_{\rho \ge 0.1 \rho_{\rm max}} r_i r_j \rho
d\mbox{\boldmath $r$}\;,
\label{eq:I}
\end{equation}
following the analysis of \citet{Matsumoto03b}. The subscripts $i$ and
$j$ represent coordinate labels, i.e., $x=r_1$, $y=r_2$, and $z=r_3$.
The three axis lengths $a_n$ are defined as the square roots of three
eigenvalues $\lambda_n$ of the specific moment of inertia $I_{ij}$, as
given by
\begin{equation}
a_n = \lambda_n^{1/2} \;,
\end{equation}
\begin{equation}
\left(
\begin{array}{ccc}
\lambda_1 & &\\ & \lambda_2 &\\ & &\lambda_3
\end{array}
\right) = \bmath{P}^{-1} \bmath{I} \bmath{P}\;,
\end{equation}
and
\begin{equation}
\bmath{P} = \left( \bmath{p}_1, \bmath{p}_2, \bmath{p}_3 \right) \;,
\end{equation}
where $\bmath{p}_n$ denote the eigenvectors of $\bmath{I}$. The axis
lengths $a_n$ correspond to the mass-weighted radii of interest. Three
eigenvectors $\bmath{p}_n$ correspond to the directions of the
principal axes of the dense region. Arranging the eigenvalues in
descending order ($\lambda_1 \ge \lambda_2 \ge \lambda_3 $), the
minimum eigenvalue $a_3$ denotes the minimum axis and corresponds to
the thickness of the disk-like cloud. The associated eigenvector
$\bmath{p}_3$ denotes the normal vector of the disk-like cloud. The
eigenvector can used for evaluating the inclination of the disk, as
shown later.

Figure~\ref{omega-B-rho.eps} shows the evolution of the central
angular velocity $\Omega_c$ defined above and the central magnetic
field strength $|\bmath{B}_c|$ (see eq.~[\ref{eq:Bc}]). It can be
clearly seen that the angular velocity and magnetic field strength
increase in proportion to $\rho_{\rm max}^{1/2}$ up to the stage of
adiabatic core formation (see lines for model MF45), implying that the
cloud undergoes self-similar collapse, maintaining its disk shape
\citep[see][]{Nakamura95,Nakamura99}. In the early accretion phase
($\rho_{\rm max} \ge \rho_{\rm cr}$), the growth of the magnetic field
and angular velocity are faster than in the runaway collapse
phase. After outflow formation ($\rho_{\rm max} \simeq
10^{-11}$~g~cm$^{-1}$), this growth slows.

As collapse proceeds, the central dense region spins up faster than
the less-dense region. The magnetic field threading the dense region
is thus wound up and twisted, causing the angular momentum to be
redistributed, and it is this process that is called magnetic
braking. In order to estimate the magnitude of magnetic braking, the
angular momentum in the dense region was followed more
closely. Figure~\ref{j_clump2.eps} shows the evolution of the central
angular momentum after decomposition into two components, parallel and
perpendicular to the magnetic field, as defined by
\begin{equation}
J_\parallel = | \bmath{J}_c \cdot \bmath{B}_c | / | \bmath{B}_c | \;,
\end{equation}
and
\begin{equation}
J_\perp = | \bmath{J}_c \times \bmath{B}_c | / | \bmath{B}_c | \;,
\end{equation}
respectively.

As collapse proceeds, $J_{\perp}/M^2$ decreases considerably, while
$J_{\parallel} / M^2$ decreases only slightly in the runaway collapse
phase. The ordinate, $J/M^2$, approximated to the specific angular
momentum per mass, $j/M$, remains constant as the disk-like cloud
collapses in the cylindrically radial direction \citep*[see,
e.g.,][]{Matsumoto97,Matsumoto99}. The decrease in both components is
therefore attributed to magnetic braking, by which angular momentum is
transferred from the central dense region to the surrounding infalling
envelope. In the range $ 10^{-14}\,{\rm g}\,{\rm cm}^{-3} \lesssim
\rho_{\rm max} \lesssim 10^{-12}\,{\rm g}\,{\rm cm}^{-3}$,
$J_{\perp}/M^2$ increases due to transfer of the angular momentum back
from the infalling envelope to the dense region. Similar undulations
of angular momentum also occurred in the perpendicular rotator of
\citet{Mouschovias79}. In the accretion phase, $J_{\perp}/M^2$
decreases drastically after outflow formation. In the outflow
formation phase, the angular momentum is therefore aligned with the
magnetic field. The parallel component $J_\parallel/M^2$ continues to
decrease considerably in this phase, similar to the behavior in
axisymmetric models \citep[e.g.,][]{Tomisaka98,Tomisaka02}.

Figure~\ref{inclination_BW.eps} shows the loci of the central magnetic
field $\bmath{B}_c$, the central angular momentum $\bmath{J}_c$, and
the normal vector of the disk-like central cloud $\bmath{p}_3$. Each
vector is plotted in the two-dimensional plane as $\arctan(V_z /
V_{xy}) V_{xy}^{-1}( V_x, V_y)$, where $\bmath{V} = (V_x, V_y, V_z) $
represents a given vector and $V_{xy} = (V_x^2+V_y^2)^{1/2}$. The
distance from the origin represents the angle between a given vector
and the $z$ axis. For example, a vector parallel to the $x$ axis is
plotted at $(90^\circ, 0^\circ)$. The direction of the magnetic field
at the boundary $r = 8 R_c$ is also plotted to represent the direction
of the global magnetic field $\bmath{B}_{\rm b}$. This magnetic field
rotates at a constant angular velocity $\Omega_0$.

The central magnetic field starts from the point $(45^\circ,
0^\circ)$, and the central angular momentum starts from the origin
according to the initial condition. The normal direction of the
density structure is located at $\sim (30^\circ, 0^\circ)$ in the very
beginning of the runaway collapse phase. At adiabatic core formation,
the central magnetic field rotates around the $z$ axis with precession
of up to $28^\circ$ in the runaway collapse phase, while the direction
of the global magnetic field rotates around the $z$ axis by only
$19^\circ$ in this phase. This difference in the direction of the
central and global magnetic fields is due to the spin-up of the dense
region. The angular momentum also undergoes precession, gradually
coming to be aligned with the central magnetic field as $J_\perp$
decreases faster than $J_\parallel$ (Fig. ~\ref{j_clump2.eps}). The
normal vector of the disk is also aligned with the central magnetic
field, suggesting that the cloud collapses along the magnetic field
lines, and the disk-like structure is formed perpendicular to the
local magnetic field. At the stage of adiabatic core formation
($\rho_{\rm max} = \rho_{cr}$), the angle between the central angular
momentum and the central magnetic field increases, due to the increase
of $J_\perp$ up to the stage of $\rho_{\rm max} \simeq
10^{-12}$~cm~g$^{-1}$ (Fig.~\ref{j_clump2.eps}). During outflow
formation, $\bmath{B}_c$, $\bmath{J}_c$, and $\bmath{p}_3$ begin to
come into alignment, and the three vectors eventually become
parallel. The central magnetic field $\bmath{B}_c$ rotates around the
$z$ axis by $\theta_z = 62^\circ$.

Figure~\ref{inclination2_BW.eps} shows the evolution of $\bmath{B}_c$,
$\bmath{J}_c$, and $\bmath{p}_3$ with respect to the global magnetic
field $\bmath{B}_{\rm b}$. In this figure, the origin represents the
direction parallel to the global magnetic field. In the early stage of
the runaway collapse phase, $\bmath{J}_c$ and $\bmath{p}_3$ pursue
$\bmath{B}_c$, and thereby move toward the origin in this
plot. However, $\bmath{B}_c$ moves to the right in this plot,
reflecting the spin-up of the central cloud. 
In the final stage, the three vectors are inclined from
the global magnetic field by $\simeq 30^\circ$. As shown later, this
angle decreases with increasing initial magnetic field strength.

Figure~\ref{radialdist.eps} shows the overall structure of the cloud
core in the final stage. Figure~\ref{radialdist.eps}{\it a} shows the
directions of the mean magnetic field, mean angular momentum, and
normal vector of the disk as a function of radius. These vectors are
evaluated in the same manner as for $\bmath{B}_c$, $\bmath{J}_c$ and
$\bmath{p}_3$, except that the volume of integration is bounded within
a given radius in equations (\ref{eq:Bc}), (\ref{eq:Jc}),
(\ref{eq:M}), and (\ref{eq:I}). The magnetic field and normal vector
of the disk are almost parallel at all radii, whereas the angular
momentum gradually becomes aligned with the other directions as $r$
decreases from $\simeq 10^4$~AU to $r \simeq 100$~AU. In $r \lesssim
100$~AU, all three vectors are roughly aligned, and the magnetic field
and normal vector coincide remarkably.

It is noteworthy that the curves in Figure~\ref{radialdist.eps}{\it a}
are quite similar to those in Figure~\ref{inclination_BW.eps} except
for the lack of precession in the former. This indicates that the
magnetic field, density distribution, and angular momentum
distribution are affected in part by the history of cloud
collapse. The early distribution is maintained in the envelope during
evolution in the central dense region, as the less-dense envelope has
a longer dynamic (freefall) timescale.

Figure~\ref{radialdist.eps}{\it b} shows the mean magnetic field
strength and rotation velocity with respect to radius $r$. The
magnetic field strength increases monotonically up to $\simeq 1$~G
toward the center, with amplification of up to $\simeq 10^5$ times. In
contrast, the rotation velocity gradually decreases from the boundary
to $10^3$~AU, but increases rapidly in the interval between $10^3$~AU
and 10~AU, reaching a maximum of $\simeq 0.2$~km~s$^{-1}$, which is
approximately equal to the sound speed $c_s$.  
The rotation curve for $r \lesssim 10$~AU reflects the
rigid rotation of the adiabatic core.
This rotatin velocity is considerably lower than the Keplerian
velocity of $\sim 2$~km~s$^{-1}$ at $\sim 10$~AU, indicating that the
adiabatic core and the surrounding disk are not rotationally
supported.  However, the gas in the outflow region spins at the
rotation velocity of $2-4$~km~s$^{-1}$, roughly equal to the outward
radial velocity of the outflow. This high-speed rotation disappears from
Figure~\ref{radialdist.eps}{\it b} in the process of averaging.

\subsubsection{Models with Different $\theta$}
\label{sec:theta}

The effect of the angle $\theta$ is investigated by comparing models
MF00, MF45, and MF90. Model MF90 is an extreme model, in which the
magnetic field is aligned perpendicular to the angular momentum in the
initial stage. The results for MF70 and MF80 are also shown for
comparison with this extreme model.

Figure~\ref{omega-B-rho.eps} shows the central magnetic field $B_c$
and the central angular velocity $\Omega_c$ as a function of maximum
density $\rho_{\rm max}$ for models MF00, MF90 and MF45. For all
models, the amplification of the magnetic field due to cloud collapse
is almost identical. The magnetic field increases in proportion to
$\rho_{\rm max}^{1/2}$ in the runaway collapse phase, irrespective of
$\theta$. This suggests that the central disk threaded perpendicularly
by the magnetic field collapses along the major axes (cylindrically
radial direction), regardless of $\theta$. However, the evolution of
the central angular velocity $\Omega_c$ does depend on $\theta$, and
models with larger $\theta$ have suppressed amplification of
$\Omega_c$. In particular, $\Omega_c$ undulates and changes sign
around the stage of $\rho_{\rm max} \sim 10^{-14}$~g~cm$^{-3}$ in
model MF90 due to strong magnetic braking. In this extreme model, the
angular momentum remains parallel to the $z$ axis due to the symmetry
assumed in the initial stage. This change in the sign of $\Omega_c$
represents anti-rotation of the central cloud
\citep[c.f.,][]{Mouschovias79}.

Figure~\ref{j_clump2.eps} shows the evolution of the central angular
momentum $\bmath{J}_c$ by decomposition into parallel and
perpendicular components with respect to the central magnetic
field. The parallel component $J_\parallel/M^2$ of model MF00 slightly
decreases in time, similar to model MF45. Model MF00 has only a
parallel component, as the perpendicular component vanishes
($J_\perp=0$) due to the axisymmetry of the model. The perpendicular
component $J_\perp/M^2$ of MF90 decreases considerably, also similar
to model MF45. After outflow formation, $J_\perp$ is expelled almost
completely from the central region and adiabatic core. Therefore, the
magnetic braking against each component is roughly independent of
$\theta$, with the perpendicular component being more strongly
affected by magnetic braking than the parallel component.

The total angular momentum of model NM, which has no magnetic field,
is also plotted for comparison in Figure~\ref{j_clump2.eps}. The
angular momentum increases slightly, roughly $\propto \rho_{\rm
max}^{1/10}$, as collapse proceeds. The slight increase in $J/M^2$ for
model NM seems to originate from the incorporation of a spherical
collapse with the disk-like collapse, resulting in an amplification of
$J/M^2$ \citep[see, e.g.,][]{Matsumoto97,Matsumoto99}. The slight
decrease in $J_\parallel /M^2$ for models MF00 and MF45 is due to weak
magnetic braking.

Figure~\ref{model4of90} shows the outflows seen in model
MF90. Surprisingly, this extreme model also forms an outflow, where
the outflow is aligned parallel to the local magnetic field and
perpendicular to the disk. The outflow has a flat shape that is
restricted to the region near the $x-y$ plane, as seen in
Figure~\ref{model4of90}{\it a}, while the angular momentum is parallel
to the $z$ direction. This outflow is quite different from that in
models with $\theta \neq 90^\circ$. In this model, the spin of the
central adiabatic core trails the threading magnetic field.
The field lines thus pinch into the region near the equatorial plane
and bend in the plane.  
Consequently, the gas is accelerated in the plane. In contrast,
outflow is associated with twisted magnetic field lines in models with
$\theta \neq 90^\circ$ (Figs.~\ref{3dmodel4.eps}{\it c} and
\ref{3dmodel4.eps}{\it d}).

Models MF70 and MF80 were constructed in order to evaluate
off-symmetry configurations such as MF90 and
MF00. Figure~\ref{inclination_model4of70.eps} shows the evolution of
$\bmath{B}_c$, $\bmath{J}_c$, and $\bmath{p}_3$ for models MF70 and
MF80. These progressions are qualitatively similar to that in model
MF45 (Fig.~\ref{inclination_BW.eps}). The normal vector $\bmath{p}_3$
is more closely aligned with the central magnetic field $\bmath{B}_c$
in these models with larger $\theta$, and these vectors still drift
with precession. The angular momentum $\bmath{J}_c$ also exhibits
precession, but the amplitude of drift is much greater than for other
two vectors. This precession of $\bmath{J}_c$ corresponds to the
undulation of $J_\perp$ during collapse. After outflow formation,
$\bmath{J}_c$ inclines to become parallel with the other vectors
$\bmath{B}_c$ and $\bmath{p}_3$. In the final stage, the three vectors
are all roughly parallel. The convergence points form an angle of
$\simeq 35^\circ$ with the global magnetic fields in both models.

\subsection{Clouds Threaded by Strong Magnetic Fields}
\label{sec:SF}

The models with strong magnetic field were initialized with a field
strength of 37.1~$\mu {\rm G}$. The evolution of the cloud cores in
these models was followed until immediately prior to outflow
formation. Extremely high Alfv\'en velocities (e.g., $430 c_s =
82~{\rm km}~{\rm s}^{-1}$ for model SF00) were achieved in the outflow
formation stage, which severely restricted the timestep of the
simulations due to the Courant condition and rendered it necessary to
halt the simulations before outflow formation.

Figure~\ref{slice_model3of45.eps} shows the central cloud in the final
stage for model SF45. The density and velocity distributions resemble
those of model MF45 (Fig.~\ref{model4of45slice.eps}{\it c}). The
density distribution reveals a disk shape oriented perpendicular to
the local magnetic field, although this is not shown explicitly in the
figure.

Figure~\ref{inclination_model3of45.eps} shows the loci of
$\bmath{B}_c$, $\bmath{J}_c$, and $\bmath{p}_3$ for model SF45. In
contrast to model MF45, precession is suppressed, and the three
vectors are fully aligned by the end of the runaway collapse phase,
indicating that $J_\perp$ is almost entirely expelled from the region
of interest, leaving only $J_\parallel$. Furthermore, the central
magnetic field forms an angle of only $5.9^\circ$ with the global
magnetic field in the final stage, indicating that the internal
magnetic field of the cloud core changes direction only slightly.

Figure~\ref{j_clump2_model3.eps} shows the evolution of the parallel
and perpendicular components of angular momentum $\bmath{J}_c/M^2$ as
a function of maximum density $\rho_{\rm max}$ for models SF00, SF45,
and SF90. The evolution of these components is also similar to that
for the moderate field models. It can be clearly seen that
$J_\parallel$ and $J_\perp$ in model SF45 progress similarly to the
respective components of SF00 and SF90, showing a strong selective
effect of magnetic braking on $J_\perp$. In all stages, $J_\parallel$
in model SF00 is $0.6 - 0.7$ of that in model MF00 (shown for
comparison), implying that the effect of magnetic braking on
$J_\parallel$ is greater under stronger magnetic fields, although the
dependence is only minor. In fact, the dependence of the strength of
the magnetic braking effect on field strength is more prominent for
$J_\perp$. For the strong field models, $J_\perp/M^2$ also undulates
during the runaway collapse, but the amplitude of fluctuation is
considerably smaller than in the moderate field models. Furthermore,
$J_\perp/M^2$ in the strong field models decreases by a factor of
$\simeq 100$ prior to adiabatic core formation, whereas that in the
moderate field models only decreases by a factor of $\simeq 10$.

\subsection{Clouds Threaded by Weak Magnetic Fields}
\label{sec:WF}

The weak field models were initialized with a magnetic field of $B_0 =
7.42 \,\mu$G.  Although the magnetic field of these models is very weak
compared to the observed fields of dense molecular cloud cores
\citep[e.g.,][]{Crutcher99}, 
these models show the dependence on the initial field strength.


Figures~\ref{model7of45_L12} and \ref{slice_model7of45} show the
central cloud ejecting an outflow in the final stage for model
WF45. The outflow has a bipolar bubble-shaped structure,
distinguishable from that in model MF45 (Figs.~\ref{3dmodel4.eps}{\it
c} and \ref{3dmodel4.eps}{\it d}). In the outflow region, a toroidal
magnetic field is prominent compared to the poloidal field, as shown
in Figure~\ref{model7of45_L12}{\it a}. This is another type of outflow
driven by the magnetic pressure gradient of the toroidal field
component \citep[I-type flow in][]{Tomisaka02}. The direction of
outflow is no longer perpendicular to the disk of a 100~AU scale.
The disk is considerably warped, and the normal direction of
the disk varies with the radius of the disk.

Figure~\ref{radial_inclination_model7.eps} shows the warp of the disk
quantitatively in terms of the normal vector of the disk $\bmath{p}_3$
with radius. A disk of 100~AU scale is inclined with respect to the
$z$ axis by $40^\circ$ toward the $y$ direction, while a disk of a
10~AU scale is inclined by $20^\circ$ in the opposite direction. The
normal direction of the disk is roughly associated with the direction
of the magnetic field, indicating that the magnetic field controls the
warp of the disk. The direction of the normal vector of the disk is
completely parallel to the magnetic field within 3~AU, corresponding
to the radius of the adiabatic core. The direction of angular momentum
also drifts in this plane, converging with that of the magnetic field
within 3~AU. The slight change in the direction of the outflow seen in
Figure~\ref{slice_model7of45} seems to be driven by the change in the
direction of the magnetic field outward from the adiabatic core.

Figure~\ref{inclination_model7of45.eps} shows the loci of the magnetic
field $\bmath{B}_c$, the normal vector of the disk $\bmath{p}_3$, and
the angular momentum $\bmath{J}_c$ for a region of $\rho \ge 0.1
\rho_{\rm max}$. The points indicating the three directions drift in
this plane, with the loci of the magnetic field and the normal vector
of the disk resembling those in
Figure~\ref{radial_inclination_model7.eps}, suggesting that the
configuration of the magnetic field and the density distribution
reflect the history of cloud collapse. In the stage of adiabatic
core formation, these directions are notably divergent. However,
After outflow formation, the directions rapidly converge.

This convergence of directions is also seen in
Figure~\ref{j_clump2_model7.eps}. For model WF45, $J_\parallel/M^2$
oscillates remarkably compared to $J_\perp/M^2$,
attributable to the obtuse angle between $\bmath{J}_c$ and
$\bmath{B}_c$ at $10^{-13}\,{\rm g}\,{\rm cm}^{-3} \lesssim \rho_{\rm
max} \lesssim 10^{-12}\,{\rm g}\,{\rm cm}^{-3}$. In the final stage,
$J_\parallel$ is considerably larger than $J_\perp$, indicating that
the angular momentum and magnetic field are aligned at this
point. Figure~\ref{j_clump2_model7.eps} shows the evolution of
$\bmath{J}_c$ in models WF00 and WF90.  Compared with model MF00,
$J_\parallel/M^2$ in model WF00 is much larger, indicating weaker
magnetic braking in the weaker field model.

\section{Discussion}
\label{sec:discussion}

\subsection{Outflow Velocity}
\label{sec:outflowvelocity}

The present simulations have shown that outflows form in the moderate
and weak field models. Figure~\ref{ofpower_plot.eps} shows the maximum
outward radial velocity in the accretion phase for all moderate and
weak field models. The evolution of the outflow velocity can be seen
to depend on the field strength. In the moderate field models, small
bumps occur at $150 ~{\rm yr} \lesssim t-t_c \lesssim 370 ~{\rm yr}$,
representing bouncing in the central adiabatic core, where $t_c$
denotes the core formation epoch. The increase in the radial velocity
in the interval $ t - t_c \gtrsim 370 $~yr corresponds to the period
of maximum outflow velocity. The outflow velocity increases
monotonically up until the final stage in models MF00 and MF45, but
increases rapidly to $6$~km~s$^{-1}$ and then gradually decreases in
model MF90. The maximum outflow velocities are roughly equal to the
spin velocity of the outflows, implying that the outflow is
accelerated by the magneto-centrifugal mechanism
\citep[c.f.,][]{Blandford82,Pudritz86}. In contrast to the moderate
field models, the outflow velocities in the weak field models exhibit
little dependence on $\theta$. The outflow velocities increase
gradually immediately after adiabatic core formation, reaching maxima
of $\simeq 2\,{\rm km}\,{\rm s}^{-1}$ in the final stage, considerably
slower than in the moderate field models.

Figure~\ref{ofpower01_vave.eps} shows the mean Alfv\'en velocity in
the outflow region. The mean Alfv\'en velocity is defined by
\begin{equation}
v_A = 
\frac{1}{V}
\int_{v_r \ge 0.1 c_s} 
\frac{\displaystyle |\bmath{B}|}{ (4 \pi \rho )^{1/2} }
d\bmath{r}\;, \label{eq:mean_va}
\end{equation}
where $V = \int_{v_r \ge 0.1 c_s }d\bmath{r} $ denotes the
volume of $v_r \ge 0.1 c_s$. 
The Alfv\'en velocities are almost identical in each field strength model.
In the moderate models, the Alfv\'en velocities increase
rapidly at $t-t_c \sim 300 - 400$~yr, synchronizing with the outflow
formation, and decrease gradually to $3-4$~km~s$^{-1}$, which is
approximately same as the outflow velocities in the final stages.
The high-speed outflow shown in Figure~\ref{3dmodel4.eps}{\it d} 
has velocity higher than the local Alfv\'en velocity, while remaining
region has sub-Alfv\'enic velocity.
For the weak field models, the Alfv\'en velocities are kept within
$1-2$~km~s$^{-1}$ after the outflow formation.  This velocity is also
comparable to the outflow velocity.
%



The present simulations give maximum outflow velocities of $\sim
6\,{\rm km}\,{\rm s}^{-1}$ for the moderate field models, and $\sim
2\,{\rm km}\,{\rm s}^{-1}$ for the weak field models. These velocities
are significantly slower than the observed outflow velocities of
$\sim$ a few $\times 100\,{\rm km}\,{\rm s}^{-1}$
\citep[e.g.,][]{Richer00,Bachiller99} considering the typical outflow
speed of optical jets. Low-velocity outflow has also been reported in
the numerical simulations of \citet{Tomisaka02,Tomisaka98} and more
recently \citet{Allen03}.

According to the steady solutions of a jet-disk system found by
\citet{Kudoh97}, the outflow velocity increases in proportion to
$v_A^{2/3}$.
%
In the strong field models, Alfv\'en speed increases up to the order
of $100$~km~s$^{-1}$, i.e., $430 c_s = 82$~km~s$^{-1}$ for model SF00,
$319 c_s = 61$~km~s$^{-1}$ for model SF45, and $375 c_s =
71$~km~s$^{-1}$ for model SF90 in the final stage, while the moderate
field models give $v_A = 3-4$~km~s$^{-1}$.
Accordingly, the high-speed outflow of optical jets ($\sim
100$~km~s$^{-1}$) are expected to occur in a cloud core with an
initial magnetic field of $\sim 40\,\mu {\rm G}$ or more.

Figure~\ref{ofpower5_fr.eps} shows the mechanical power of the outflow
in terms of the radial momentum estimated by
\begin{equation}
P = \int_{v_r \ge 5 c_s}\rho v_r d\bmath{r}\;,
\end{equation}
where $v_r$ denotes the radial velocity.
%
Despite the slow outflow velocities in the weak field models (WF00 and
WF45), the powers of the outflows are comparable to those of the
moderate field models, implying that the volume of the outflow is also
comparable to that of the moderate field models. The power is equal to
$\simeq 10^{-4}\,M_\odot\,{\rm km}\,{\rm s}^{-1}$ in the final stage
for model MF45, and appears to continue to increase. This increase in
outflow power represents an increase in the volume or mass of the
outflow region rather than an increase in the maximum velocity. The
momentum flux is therefore $P/(t-t_c) \simeq
1.6\times10^{-7}\,M_\odot\,{\rm km}\,{\rm s}^{-1}\,{\rm yr}^{-1}$ in
the final stage for model MF45. However, this momentum flux is
considerably smaller than the observed flux of $\gtrsim
10^{-6}\,M_\odot\,{\rm km}\,{\rm s}^{-1}\,{\rm yr}^{-1}$ for the
bipolar molecular outflows around YSOs \citep[see
e.g.,][]{Bontemps96}. \citet{Tomisaka02} has shown that the momentum
flux reaches $10^{-6}-10^{-5}\,M_\odot\,{\rm km}\,{\rm s}^{-1}\,{\rm
yr}^{-1}$ at $t-t_c \simeq 10^4$~yr, which represents much longer term
evolution compared to the simulations in this study.

We estimate the mass flux in the outflow ($\dot{M}_{\rm out}$) 
and the accretion rate ($\dot{M}_{\rm in}$)
across the boundary of the 13th level grid, of which box-size is 71~AU, at
the final stages.
The outflow rates are 
$\dot{M}_{\rm out}
= 6.56\times10^{-7}\,M_\odot\,{\rm yr}^{-1}$,
$4.97\times10^{-7}\,M_\odot\,{\rm yr}^{-1}$, and
$8.01\times10^{-7}\,M_\odot\,{\rm yr}^{-1}$
for models MF00, MF45, and MF90 while the accretion rates are 
identical to each other ($\dot{M}_{\rm in} = 4.1\times10^{-5}\,M_\odot\,{\rm yr}^{-1}$).
A model with larger outflow rate shows larger outflow power $P$ at the
final stages. 
The ratio $\dot{M}_{\rm out} /\dot{M}_{\rm in}$ ranges from 1.2\% to 2.0\%.
The outflow rates in the weak field models are larger than those in
the moderate field models such that
$\dot{M}_{\rm out} 
=4.45\times10^{-6}\,M_\odot\,{\rm yr}^{-1}$,
$3.90\times10^{-6}\,M_\odot\,{\rm yr}^{-1}$, and
$1.74\times10^{-6}\,M_\odot\,{\rm yr}^{-1}$
for models WF00, WF45, and WF90.
These are also in the same order of strength as the outflow power.
The accretion rates are
$\dot{M}_{\rm in} = (4.32-5.62)\times10^{-5}\,M_\odot\,{\rm yr}^{-1}$,
which are approximately equal to that in the moderate field models.
The ratio $\dot{M}_{\rm out} /\dot{M}_{\rm in}$ ranges from 3.1\% to 10\%.
The ratios of $\dot{M}_{\rm out} /\dot{M}_{\rm in}$ presented here are
considerably smaller than that of \citet{Tomisaka02} because of 
the early phase of outflow.


\subsection{Precession and Convergence of Magnetic Field, Angular Momentum, and Disk}
\label{sec:precession}

The cloud collapses with precession of the central magnetic field,
angular momentum, and shape of the disk such that in the final stage,
the directions of these descriptors converge to a single direction as
the precession decreases in amplitude, as seen in
Figures~\ref{inclination_BW.eps}, \ref{inclination_model4of70.eps},
and \ref{inclination_model7of45.eps}. The precession and convergence
of the magnetic field and angular momentum is due to the interaction
between them.  Figure~\ref{schematic.eps} shows a schematic diagram of
this interaction, which consists of two mechanisms; rotation of the
magnetic field and magnetic braking. The magnetic field threading the
central cloud rotates due to the spin of the cloud, that is, the
rotation axis of the magnetic field is aligned parallel to the angular
momentum (Fig.~\ref{schematic.eps}{\it a}). The direction of the
magnetic field $\bmath{B}_c$ drifts around a guiding center
represented by the direction of the angular momentum $\bmath{J}_c$, as
seen in Figures~\ref{inclination_BW.eps},
\ref{inclination_model4of70.eps}, and
\ref{inclination_model7of45.eps}. The angular momentum is subsequently
affected by magnetic braking. In the runaway collapse phase, $J_\perp$
undulates as shown above \citep[c.f.,][]{Mouschovias79}, which also
causes the direction of angular momentum to oscillate
(Fig.~\ref{schematic.eps}{\it b}). In the outflow formation phase, the
angular momentum rapidly inclines toward the local magnetic field,
that is, the guiding center moves due to $\bmath{B}_c$, and the
direction of $\bmath{J}_c$ follows, as shown in
Figures~\ref{inclination_BW.eps}, \ref{inclination_model4of70.eps},
and \ref{inclination_model7of45.eps}. Consequently, the magnetic field
and angular momentum precess until they come into alignment.

The magnetic braking timescale can be estimated by a similar method to
that employed in \citet{Mouschovias80} and \citet{Dorfi89}. The
Alfv\'en wave transfers angular momentum from the central dense region
to the outer less-dense region. Although the dense region of the
collapsing magnetized cloud has a disk shape in the simulations, the
central region of interest, i.e., the region of $\rho \ge 0.1
\rho_{\rm max}$, is less oblate.  Thus, the density distribution of
the central collapsing cloud can be approximated by a spherical cloud
for simplicity, as given by
\begin{equation}
\rho \simeq \left\{
\begin{array}{ll}
\rho_c & {\rm for}\; r < \lambda_J\\ \rho_c (r/\lambda_J)^{-2} & {\rm
for}\; r \ge \lambda_J
\end{array}
\right. \;,
\end{equation}
where $\rho_c$ denotes the central density and $\lambda_J = c_s(\pi/G
\rho_c)^{1/2}$ denotes the Jeans length. The cloud has a central
plateau and surrounding envelope, where the plateau maintains its
scale length to be roughly equal to the Jeans length, which decreases
in proportion to $\rho_c^{-1/2}$. The $z$ component of the moment of
inertia of the plateau is estimated as
\begin{eqnarray}
I_c &=& \int_{ r < \lambda_J} \rho(x^2+y^2) d\bmath{r} \\ &=&
    \frac{8}{15} \pi \rho_c \lambda_J^5\;.
\end{eqnarray}
Similarly, the moment of inertia of the envelope where the Alfv\'en
wave sweeps is estimated as
\begin{eqnarray}
I_{\rm ext} (t) &=& \int_{ \lambda_J \le r \le v_a t} \rho(x^2+y^2)
               d\bmath{r}\\ &=& \frac{8}{9} \pi \rho_c \lambda_J^5
               \left[\left(\frac{v_a t}{\lambda_J}\right)^3 -1\right]
               \;,
\end{eqnarray}
where $v_a$ denotes the Alfv\'en velocity. The braking time $t_b$ is
estimated by equating the moments of inertia,
\begin{equation}
I_c = I_{\rm ext}(t_b)\;,
\end{equation}
leading to the braking time of
\begin{equation}
t_b = \left( \frac{8}{5} \right)^{1/3} \frac{\lambda_J}{v_a}\;.
\label{eq:tb}
\end{equation}
The Alfv\'en velocity $v_a$ remains roughly constant because the
central magnetic field increases in proportion to $\rho_{\rm
max}^{1/2}$ (see Fig.~\ref{omega-B-rho.eps}). However, the Jeans
length $\lambda_J$ decreases in proportion to $\rho_{\rm max}^{-1/2}$
in the runaway collapse phase, with the result that the braking time
decreases during the collapse in proportion to $\rho_{\rm
max}^{-1/2}$. It should be noticed that the braking time has the same
dependence $\rho_{\rm max}^{-1/2}$ as that of the freefall timescale
and the rotational period (without magnetic braking).

The infalling envelope shows an infall velocity higher than the
Alfv\'en velocity in the runaway collapse phase, and one might image
that the Alfv\'en wave is trapped in the region of the central density
plateau. The plateau shrinks, and its boundary moves inward at the
speed higher than that of the outward propagating Alfv\'en wave ($v_r +
v_a$).  This indicates that the Alfv\'en wave can escape from the
central plateau and it transfers angular momentum to the infalling
envelope.  In the accretion phase, the Alfv\'en velocity around the
adiabatic core becomes extremely high ($\sim 100\,{\rm km}\,{\rm
s}^{-1}$), and the infalling gas has sub-Alfv\'enic velocity.

Although the precession and convergence are attributed to nonlinear
interaction between magnetic braking and rotation, the magnetic
braking during collapse can be roughly estimated by applying equation
(\ref{eq:tb}). Given the initial state of the present simulations, the
braking times are $t_b = 1.64 \times 10^5$~yr, $3.28 \times 10^6$~yr,
and $8.20 \times 10^6$~yr for the strong, moderate and weak field
models, respectively.  The precession is also related to the other
timescales of the freefall time $t_{\rm ff}$, and rotation timescale
$t_{\rm rot}\, (\equiv 1/\Omega_c)$: $t_{\rm ff} = 2.10\times10^5$~yr
and $t_{\rm rot} = 1.28\times10^6$~yr in all models, where the
rotation time is 6.6 times longer than the freefall time. In the
strong field models, the braking time is considerably shorter than the
freefall time, which allows magnetic braking to suppress the spin-up
of the dense region during the collapse, as seen in Figure
\ref{inclination_model3of45.eps}. In the moderate field models, the
braking time is longer than the freefall time, resulting in
considerable precession during the collapse as shown in Figures
\ref{inclination_BW.eps} and \ref{inclination_model4of70.eps}. In the
weak field models, the braking time is much longer than the freefall
time, and the magnetic braking is too weak to redistribute the angular
momentum significantly during the collapse. Moreover, the rotation
timescale is comparable to the braking time, resulting in large
precession, as shown in Figure \ref{inclination_model7of45.eps}.



The precession of the disk axis is attributed to the precession of the
magnetic field, which controls the collapse of the cloud by guiding
the infalling gas. As the disk forms perpendicular to the direction of
gas infall, the disk axis follows the direction of the magnetic field
during the collapse. In this sense, the precession of the disk axis is
passive.


Although the cloud core is affected by magnetic braking in the
simulations, the dense adiabatic core (the seed of a protostar) may be
decoupled from the magnetic field as a result of efficient ambipolar
diffusion \citep*{Nakano02} in further stages. The directions of the
disk, rotation axis, and outflow will be fixed at the convergence
direction during the main accretion phase of the protostar.

\subsection{Magnetic Braking and Torque}
\label{sec:torque}

The torque working on the dense region of $\rho \ge 0.1\rho_{\rm max}$
during the collapse is estimated by performing $\int_{\rho \ge
0.1\rho_{\rm max}} \bmath{r} \times (\cdot) d \bmath{r}$ for
equation~(\ref{eq:momentum}), which yields
\begin{equation}
\frac{d\bmath{J}_c}{dt} = \bmath{N}_{\rm tot} \;,
\end{equation}
and
\begin{equation}
\bmath{N}_{\rm tot} = \bmath{N}_P + \bmath{N}_m + \bmath{N}_t +
\bmath{N}_g + \bmath{N}_a \;.
\end{equation}
The torques due to thermal pressure ($N_P$), magnetic pressure
($N_m$), magnetic tension ($N_t$), and gravity ($N_g$), are then
defined as
\begin{eqnarray}
\bmath{N}_P &=& -\int_{\rho \ge 0.1\rho_{\rm max}} \bmath{r} \times
\nabla P d \bmath{r}\;,\\ \bmath{N}_m &=& -\int_{\rho \ge 0.1\rho_{\rm
max}} \bmath{r} \times \nabla \left(\frac{|\bmath{B}|^2}{8\pi}\right)
d \bmath{r}\;,\\ \bmath{N}_t &=& \frac{1}{4\pi}\int_{\rho =
0.1\rho_{\rm max}} (\bmath{r}\times\bmath{B}) (\bmath{B}\cdot
d\bmath{S})\;,\\ \bmath{N}_g &=& -\int_{\rho \ge 0.1\rho_{\rm max}}
\rho \bmath{r} \times \nabla\Psi d \bmath{r}\;,
\end{eqnarray}
where $\int d\bmath{S}$ denotes the surface integral. The advection of
angular momentum ($N_a$) is defined as
\begin{equation}
\bmath{N}_a = - \int_{\rho = 0.1\rho_{\rm max}} \rho
(\bmath{r}\times\bmath{v}) ( \bmath{v}\cdot d\bmath{S} ) \;.
\end{equation}
It should be pointed out that the volume of the above integral (region
of $\rho \ge 0.1\rho_{\rm max}$) changes temporally. In other words,
the region of interest containing $\bmath{J}_c$ shrinks during the
collapse. Accordingly, $\bmath{N}_{\rm tot}$ does not indicate an
increasing rate of $\bmath{J}_c$ in its temporal evolution. However,
it is true that $\bmath{N}_{\rm tot}$ represents the torque
functioning on the region of interest, $\rho \ge 0.1\rho_{\rm max}$.

Figure \ref{clump_torque.eps}{\it a} shows the torques per mass
$|\bmath{N}|/M$ as a function of $\rho_{\rm max}$ for model MF45. The
torque of magnetic tension $N_t$ and the advection of angular momentum
$N_a$ are much larger than the other components. The torque of
magnetic tension $N_t$ causes magnetic braking, while the advection of
angular momentum $N_a$ drives the spin-up of the central cloud. In the
runaway collapse phase, $N_t/M$ remains almost constant, as indicated
by equation (\ref{eq:tb}) and the relationship of $M \propto \rho_{\rm
max}^{-1/2}$. In the accretion phase, $N_t/M$ increases, promoting
strong magnetic braking. Similarly, $N_a/M$ remains constant in the
runaway collapse phase because $v_R \propto R$, $j \propto R$, $\rho
\propto R^{-2}$, and $M \propto R$. Therefore, $M^{-1} \int \rho j v_R
R dR $ is constant, where $R$ and $j$ denote the cylindrical radius
and the specific angular momentum.

Figure \ref{clump_torque.eps}{\it b} shows the parallel and
perpendicular components of $\bmath{N}_t/M$, $\bmath{N}_a/M$, and
$\bmath{N}_{\rm tot}/M$ with respect to the angular momentum
$\bmath{J}_c$ as a function of $\rho_{\rm max}$. The parallel and
perpendicular components are defined as
\begin{equation}
N_\parallel = |\bmath{N} \cdot \bmath{J}_c|/|\bmath{J}_c| \;,
\end{equation}
and
\begin{equation}
N_\perp = |\bmath{N} \times \bmath{J}_c|/|\bmath{J}_c| \;.
\end{equation}
The perpendicular component changes the direction of $\bmath{J}_c$. In
the runaway collapse phase, $N_{a, \perp}$ is much smaller than $N_{a,
\parallel}$, indicating that gas accretion does not change the
direction of rotation considerably. On the other hand, $N_{t,\perp}$
remains roughly constant while $N_{t, \parallel}$ oscillates in the
runaway collapse phase. In $10^{-16}\,{\rm g}\,{\rm cm}^{-3} \lesssim
\rho_{\rm max} \lesssim 10^{-12}\,{\rm g}\,{\rm cm}^{-3}$, $N_{t,
\perp}$ is larger than $N_{t,\parallel}$, with the result that the
tension of the magnetic field changes the direction of $\bmath{J}_c$
considerably.
%

In the early accretion phase, $N_{{\rm tot},\perp}$ is considerably
larger than $N_{{\rm tot},\parallel}$ due to large $N_{t,\perp}$. This
torque drives $\bmath{J}_c$ to come into alignment with $\bmath{B}_c$
in the outflow formation epoch, resulting in the convergence of
$\bmath{J}_c$ as shown in Figure \ref{inclination_BW.eps}.

\subsection{Application to Observations}
\label{sec:alignment}

Observations of optical polarization reveal that the optical outflows
tend to be aligned with the cloud-scale magnetic
field. \citet{Cohen84} showed that half of the well-defined outflows
are aligned with the magnetic field within $\pm 20^\circ$, and the
remainder except for only one outflow are aligned within $\pm 40^\circ$ 
in the Cepheus A
region. Similarly, \citet{Vrba88} showed that the position angles of
the outflows are inclined by $40^\circ$ from the cloud-scale magnetic
field in the Orion region. The results of the present simulations are
highly consistent with these observations.

The magnetic field lines inside the molecular cloud deviate
significantly, as seen in Figures \ref{radialdist.eps}{\it a} and
\ref{radial_inclination_model7.eps}, and when the outflow extends out
to $\gtrsim 100$~AU, it is no longer parallel with the local magnetic
field.  The outflow velocity is comparable to the
Alfv\'en velocity, as seen in Figures \ref{ofpower_plot.eps} and
\ref{ofpower01_vave.eps}. As the field strength is much weaker at the
outer radius of the molecular cloud core, as seen in Figure
\ref{radialdist.eps}{\it b}, the outflow can propagate straight in the
direction in which the gas is launch at the root of the
outflow. Accordingly, the outflow controls the direction of the
magnetic field at the $\gtrsim$ 100~AU scale of the cloud core, as
observed by \citet{Momose01}, although they argued that the magnetic
field controls the outflow direction.

At the cloud-core scale, the configuration of the magnetic field is
more complicated, as suggested by observations of submillimeter
polarization \citep[e.g.,][]{Wolf03}. In order to compare the present
results with those observations, it is necessary to reproduce the
polarization map using the simulation data. Such work is planned as
part of future study.

\section{Summary}
\label{sec:summary}

The gravitational collapse of slowly rotating, magnetized molecular
cloud cores was investigated by high-resolution numerical simulations.
Outflow formation in the collapsing cloud cores was also followed in
the simulations. The cloud was found to collapse along magnetic field
lines to form a disk perpendicular to the magnetic field. The central
dense region spins up as a consequence of collapse. This spin-up is
counteracted by magnetic braking, which was found to be more effective
against the perpendicular component of angular momentum with respect
to the local magnetic field than the parallel component. This leads to
alignment of the magnetic field and rotation axis. With an initial
magnetic field of $\sim 40\,\mu {\rm G}$, the magnetic field and
rotation axis converge to alignment prior to the formation of an
adiabatic core. With weaker initial magnetic fields, convergence also
occurs rapidly due to the torque of magnetic tension associated with
outflow formation. The central magnetic field is inclined from the
cloud-scale magnetic field by no more than $\sim \, 30^\circ$ in the
outflow formation stage given an initial magnetic field of $\sim \,
20\,\mu {\rm G}$, and the inclination decreases with increasing
initial field strength.

The history of cloud collapse reproduces the distribution of
magnetic field and density in the cloud core. The formation of a flat
disk-like envelope at every scale of the cloud core occurs due to the
perpendicular alignment of the disk with respect to the local magnetic
field. In the envelope, the angular momentum is not required to be
parallel to the local magnetic field, but becomes parallel at small
scales of $10$~AU.

The outflow is always parallel to the magnetic field at the root of
the outflow and perpendicular to the disk. In models with $\theta
\lesssim 80^\circ$, the central cloud converges to a near-axisymmetric
model, and the outflow is attributed to the magneto-centrifugal
wind. Even for models with $\theta = 90^\circ$, the outflow is aligned
parallel to the local magnetic field, and therefore perpendicular to
the rotation axis. The acceleration is attributed to the pinch of the
magnetic field.

Consequently, disk-outflow systems are aligned with the cloud-scale
magnetic field within $\sim 30^\circ$ and $\sim 5^\circ$ for initial
field strengths of $\sim20\,\mu{\rm G}$ and $\sim40\,\mu{\rm G}$. This
result is highly consistent with observations of the optical
polarization of Cepheus A and Orion.

\acknowledgments

The authors thank M.~Tamura for valuable discussion. Numerical
computations were carried out on the VPP5000 supercomputer at the
Astronomical Data Analysis Center of the National Astronomical
Observatory, Japan, which is an inter-university astronomy research
institute operated by the Ministry of Education, Science, Sports,
Culture and Technology, Japan. This research was supported in part by
Grants-in-Aid for Young Scientists (B) 14740134 and 16740115 (TM), for
Scientific Research (C) 14540233 (KT), and for Scientific Research (B)
13011204 (TM, KT) by the Ministry of Education, Science, Sports,
Culture and Technology, Japan




\clearpage

\begin{figure}[p]
\epsscale{0.9} 
\figcaption[model4of45slice.eps]{ Evolution of moderate field model
MF45. Density is shown in grayscale, and radial velocity is shown as
arrows and red (positive) and blue (negative) contours at
0.2~km~s$^{-1}$ intervals. The grid level is indicated by \#$l$.
({\it a-c}) Cross-sections taken along plane A at various time points,
shown at different scales. ({\it d,e}) Cross-sections along ({\it d})
plane A and ({\it e}) plane B at the same time point. ({\it f})
Schematic showing planes A and B.
%
\label{model4of45slice.eps}
}
\end{figure}

\begin{figure}[p]
\epsscale{0.9} 
\figcaption[3dmodel4.eps]{ Three-dimensional structures of density and
magnetic field in the final stage of model MF45. Isosurfaces denote
the iso-density surfaces of ({\it a}) $\rho = \rho_0$, ({\it b}) $10^2
\rho_0$, and ({\it c}) $10^5 \rho_0$. Stream lines denote magnetic
field lines, and the grid level is indicated by \#$l$. The box sizes
are ({\it a}) 0.356~pc, ({\it b}) 4590~AU, and ({\it c,d}) 287~AU,
where ({\it d}) shows the positive radial velocity (outflows) as $v_r
= 10 c_s$ iso-velocity surfaces.
\label{3dmodel4.eps}
}
\end{figure}

\begin{figure}[p]
\epsscale{0.80} 
\caption{ Magnetic field strength and angular velocity in the dense
region ($\rho \ge 0.1 \rho_{\rm max}$) as a function of maximum
density $\rho_{\rm max}$ for models MF00, MF45, and MF90. Upper solid
lines denote magnetic field strength $B_c (c_s^2 \rho_0)^{-1/2}$, the
dotted line denotes the relationship $\propto \rho_{\rm max}^{1/2}$,
the dashed line denotes angular velocity of model NM, lower solid
lines denote angular velocity $\Omega_c (4 \pi G \rho_0)^{-1/2}$, and
diamonds denote the stage of adiabatic core formation ($\rho_{\rm
max}= \rho_{cr}$).
\label{omega-B-rho.eps}
}
\end{figure}


\begin{figure}[p]
\epsscale{0.80} 
\caption{ Angular momentum as a function of maximum density for models
MF00, MF45, and MF90. Dashed line denotes $J_c/M^2 (c_s/4 \pi G)$ for
model NM, solid lines denote angular momentum parallel to the local
magnetic field $J_{\parallel}/M^2 (c_s/4 \pi G)$, dotted lines denote
angular momentum perpendicular to the local magnetic field
$J_{\perp}/M^2 (c_s/4 \pi G)$ and diamonds denote the stage of
adiabatic core formation ($\rho_{\rm max}= \rho_{cr}$).
\label{j_clump2.eps}
}
\end{figure}

\begin{figure}[p]
\epsscale{0.80} 
\caption{ Loci of directions of the central magnetic field
$\bmath{B}_c$ (thin solid line), the central angular momentum
$\bmath{J}_c$ (dotted line), the normal vector of the disk-like
density structure $\bmath{p}_3$ (dashed line), and the magnetic field
at the boundary $\bmath{B}_{\rm b}$ (thick line) for model
MF45. Diamonds denote the stage of adiabatic core formation
($\rho_{\rm max} = \rho_{cr}$), and triangles denote the final
stage. ({\it Inset}) Detail of the convergence point.
\label{inclination_BW.eps}
}
\end{figure}

\begin{figure}[p]
\epsscale{0.80} 
\caption{ Loci of directions of the central magnetic field
$\bmath{B}_c$ (thin solid line), the central angular momentum
$\bmath{J}_c$ (dotted line), and the normal vector of the disk-like
density structure $\bmath{p}_3$ (dashed line) with respect to the
magnetic field at the boundary $\bmath{B}_{\rm b}$ for
model MF45. Diamonds denote the stage of adiabatic core formation
($\rho_{\rm max} = \rho_{cr}$), and triangles denote the final stage.
\label{inclination2_BW.eps}
}
\end{figure}


\begin{figure}[p]
\epsscale{0.65} 
\caption{ Radial distribution of the mean magnetic field, mean angular
momentum, and normal vector of the disk in the final stage of model
MF45. ({\it a}) Directions of the magnetic field (solid line), angular
momentum (dotted line), and normal vector of the disk (dashed line)
for various radii. ({\it b}) Strength of mean magnetic field (solid line)
and mean rotation velocity (dotted line) as a function of radius.
\label{radialdist.eps}
}
\end{figure}


\begin{figure}[p]
\epsscale{0.45} 
\caption{ Density, velocity and magnetic field distributions for model
MF90. ({\it a}) Three-dimensional structure of the outflow region at $t-t_c
= 464$~yr, where the outflow speed reaches its maximum of
$6.14$~km~s$^{-1}$. Streamlines denote the magnetic field lines, and
the disk-like surface represents the iso-density surface of $10^5
\rho_0$. Other surfaces denote the iso-velocity surfaces of $v_r =
c_s$. Box size is 287~AU ($l=12$). ({\it b}) Cross-section of outflow region
in the final stage using the same notation as
Fig.~\ref{model4of45slice.eps}.
\label{model4of90}
}
\end{figure}


\begin{figure}[p]
\epsscale{0.45} 
\caption{ Loci of directions of the central magnetic field
$\bmath{B}_c$ (thin solid line), the central angular momentum
$\bmath{J}_c$ (dotted line), the normal vector of the disk-like
density structure $\bmath{p}_3$ (dashed line), and the magnetic field
at the boundary $\bmath{B}_{\rm b}$ (thick line) for models ({\it a}) MF70
and ({\it b}) MF80. Diamonds denote the stage of adiabatic core formation
($\rho_{\rm max} = \rho_{cr}$), and triangles denote the final stage.
\label{inclination_model4of70.eps}
}
\end{figure}

\begin{figure}[p]
\epsscale{0.8} 
\caption{ Structure of strong field model SF45 in the final
stage. Density is shown in grayscale, and radial velocity is shown as
arrows and red (positive) and blue (negative) contours at
0.2~km~s$^{-1}$ intervals. The grid level is indicated by \#$l$. 
Cross-section is taken along plane B in
Fig.~\ref{model4of45slice.eps}{\it f}.
\label{slice_model3of45.eps}
}
\end{figure}

\begin{figure}[p]
\epsscale{0.80} 
\caption{ Loci of directions of the central magnetic field
$\bmath{B}_c$ (thin solid line), the central angular momentum
$\bmath{J}_c$ (dotted line), the normal vector of the disk-like
density structure $\bmath{p}_3$ (dashed line), and the magnetic field
at the boundary $\bmath{B}_{\rm b}$ (thick line) for model
SF45. Diamonds denote the stage of adiabatic core formation
($\rho_{\rm max} = \rho_{cr}$), and triangles denote the final stage.
\label{inclination_model3of45.eps}
}
\end{figure}

\begin{figure}[p]
\epsscale{0.80} 
\caption{ Angular momentum as a function of maximum density for models
SF00, SF45, and SF90. Dashed line denotes $J_c/M^2 (c_s/4 \pi G)$ for
model NM, solid lines denote angular momentum parallel to the local
magnetic field $J_{\parallel}/M^2 (c_s/4 \pi G)$, dotted lines denote
angular momentum perpendicular to the local magnetic field
$J_{\perp}/M^2 (c_s/4 \pi G)$, dot-dashed line denotes
$J_{\parallel}/M^2$ for model MF00, and diamonds denote the stage of
adiabatic core formation ($\rho_{\rm max}= \rho_{cr}$).
\label{j_clump2_model3.eps}
}
\end{figure}


\clearpage

\begin{figure}[p]
\epsscale{0.45} 
\caption{ Three-dimensional structure of the outflow region in the
final stage for model WF45. Box size is 287~AU ($l=12$).
({\it a}) Streamlines denote the magnetic field lines, and the disk-like surface
represents the iso-density surface of $10^5\,\rho_0$. ({\it b}) Iso-velocity
surface of $v_r = c_s$.
\label{model7of45_L12}
}
\end{figure}

\begin{figure}[p]
\epsscale{0.8}
\caption{ Structure of strong field model WF45 in the final
stage. Density is shown in grayscale, and radial velocity is shown as
arrows and red (positive) and blue (negative) contours at
0.2~km~s$^{-1}$ intervals. The grid level is indicated by \#$l$. 
Cross-section is taken along plane A in
Fig.~\ref{model4of45slice.eps}{\it f}.
\label{slice_model7of45}
}
\end{figure}

\begin{figure}[p]
\epsscale{0.80} 
\caption{ Radial distribution of the mean magnetic field (solid line),
mean angular momentum (dotted line), and normal vector of the disk
(dashed line) in the final stage of model WF45.
\label{radial_inclination_model7.eps}
}
\end{figure}

\begin{figure}[p]
\epsscale{0.80} 
\caption{ Loci of directions of the central magnetic field
$\bmath{B}_c$ (thin solid line), the central angular momentum
$\bmath{J}_c$ (dotted line), the normal vector of the disk-like
density structure $\bmath{p}_3$ (dashed line), and the magnetic field
at the boundary $\bmath{B}_{\rm b}$ (thick line) for model
WF45. Diamonds denote the stage of adiabatic core formation
($\rho_{\rm max} = \rho_{cr}$), and triangles denote the final stage.
\label{inclination_model7of45.eps}
}
\end{figure}

\begin{figure}[p]
\epsscale{0.80} 
\caption{ Angular momentum as a function of maximum density for models
WF00, WF45, and WF90. Dashed line denotes $J_c/M^2 (c_s/4 \pi G)$ for
model NM, solid lines denote angular momentum parallel to the local
magnetic field $J_{\parallel}/M^2 (c_s/4 \pi G)$, dotted lines denote
angular momentum perpendicular to the local magnetic field
$J_{\perp}/M^2 (c_s/4 \pi G)$, dot-dashed line denotes
$J_{\parallel}/M^2$ for model MF00, and diamonds denote the stage of
adiabatic core formation ($\rho_{\rm max}= \rho_{cr}$).
\label{j_clump2_model7.eps}
}
\end{figure}

\begin{figure}[p]
\epsscale{0.80} 
\caption{ Maximum radial velocities of outflows as a function of time
from the adiabatic core formation epoch for models MF00, MF45, MF90,
WF00, WF45, and WF90.
\label{ofpower_plot.eps}
}
\end{figure}

\begin{figure}[p]
\epsscale{0.80} 
\caption{ Mean Alfv\'en speeds in the outflow region ($v_r \ge 0.1
c_s$) as a function of time from the adiabatic core formation epoch
for models MF00, MF45, MF90, WF00, WF45, and WF90.
\label{ofpower01_vave.eps}
}
\end{figure}

\begin{figure}[p]
\epsscale{0.80} 
\caption{ Radial momentum of the outflow region ($v_r \ge 5 c_s$) as a
function of time from the adiabatic core formation epoch for models
MF00, MF45, MF90, WF00, WF45, and WF90.
\label{ofpower5_fr.eps}
}
\end{figure}


\begin{figure}[p]
\epsscale{0.8} 
\caption{ Schematic diagram of interaction between magnetic field and
angular momentum. ({\it a}) The magnetic field rotates due to the spin of
the cloud around the angular momentum vector. ({\it b}) The angular momentum
inclines from the magnetic field and oscillates due to magnetic
braking.
\label{schematic.eps}
}
\end{figure}

\begin{figure}[p]
\epsscale{0.65} 
\caption{ Torque per mass $\bmath{N}/M$ acting on the dense region of
$\rho \ge 0.1\rho_{\rm max}$ as a function of $\rho_{\rm max}$ for
model MF45. ({\it a}) Solid lines denote $N_t/M$, $N_a/M$, $N_m/M$, $N_g/M$,
and $N_P/M$, dashed line denotes $N_{\rm tot}/M$. ({\it b}) Thick, thin, and
gray lines denote $\bmath{N}_t/M$, $\bmath{N}_a/M$, and
$\bmath{N}_{\rm tot}/M$. Solid and dashed lines denote parallel and
perpendicular components with respect to angular momentum
$\bmath{J}_c$.
\label{clump_torque.eps}
}
\end{figure}

\clearpage

\begin{deluxetable}{lllll}
\tabletypesize{\scriptsize}
\tablecaption{Models and magnetic field strengths\label{table:mag}}
\tablewidth{0pt}
\tablehead{
\colhead{Models} & 
\colhead{$B_0$ ($\mu {\rm G}$)} & 
\colhead{$B_0/B_{\rm cr}$} &
\colhead{$E_{\rm mag}/| E_{\rm grav} |$} & 
\colhead{$\beta_p$} 
}
\startdata
SF00, SF45, SF90 & 37.1 & 0.50 & 2.88 & 0.658 \\
MF00, MF45, MF70, MF80, MF90 & 18.6 & 0.25 & 0.721 & 2.63 \\
WF00, WF45, WF90 &7.42 & 0.10 & 0.115 & 16.5 \\
NM               & 0.00 & 0.00 & 0.00 & $\infty$ \\
 \enddata
\end{deluxetable}


\end{document}